\documentclass[a4paper,11pt]{article}
\usepackage{jinstpub} 
\usepackage{lineno}
\usepackage{algorithm}
\usepackage{algpseudocode}
\usepackage{sidecap}

\usepackage{subcaption}

\title{Challenges in the verification of MOSAIX ASIC for ALICE ITS3}

\author[a,1]{S.~Scarf\`i,\note{on behalf of the ALICE collaboration}}
\author[a]{M.~Lupi,}
\author[a]{A.~Pulli,}
 
\affiliation[a]{CERN,\\
1, Esplanade des Particules, Meyrin, Switzerland}

\emailAdd{simone.scarfi@cern.ch}

\abstract{
The MOSAIX is a wafer-scale, full-functionality prototype ASIC for the ALICE Inner Tracking System (ITS3) upgrade, implemented in $65\,\text{nm}$ stitching technology. 
It integrates 144 independently powered pixel matrices, eight $10\,\text{Gb/s}$ serializers, and on-chip data and power distribution, making full-chip verification highly resource-intensive (over $80\,\text{GB}$ of memory and 192,000 tests). Verification also targets a sophisticated power-up sequence and radiation conditions of $10^{6}$\ particles $s^{-1} cm^{-2}$ and $4\,\text{kGy}$ TID. This work presents a UVM-based verification strategy that identified 260 design errors and summarizes key lessons from ASIC sign-off.
}


\arxivnumber{1111.00000}

\begin{document}
\maketitle
\flushbottom

\section{Introduction}\label{sec:intro}
In traditional detector systems, such as ALICE Inner Tracking System (ITS2), tracking layers are constructed by mounting individual readout chips onto circuit boards that host flexible cables and external components that traverse the active area to deliver power and route data, significantly increasing the material budget. To overcome these structural, electrical, and manufacturing challenges, the ALICE ITS3 upgrade will replace the inner barrel of ITS2 with ultra-thin, curved wafer-scale silicon sensors~\cite{tdr:alice:its3}. 

The MOSAIX (MOnolithic Stitched Active pIXel) ASIC was designed as a single, continuous, wafer-scale device measuring approximately $26.66\,\text{cm} \times 1.96\,\text{cm}$. By integrating power distribution, control logic, and data aggregation directly onto the silicon wafer, the MOSAIX architecture eliminates the need for external interconnects in the active region. This approach reduces the detector's material budget by $75\%$ down to just $0.09\% X_0$, thereby preventing particles from scattering~\cite{tdr:alice:its3}.

Verifying a wafer-scale ASIC of this size presents unique challenges distinct from standard chip verification. The design effectively functions as a full system comprising a high number of independent power domains, making full-chip simulation highly resource-intensive, with over $80\,\text{GB}$ of memory required. Simulation time is also long, with over one day of simulation per single run. Assuming 64 licence cores and a total of 192,000 tests (number taken from the actual ASIC signoff) for final sign-off, it would have taken around 3,000 days. For this reason, the verification problem was split into a multi-scale environment to verify certain features in a sub-environment and other features at a higher-level environment.
To address this, all testbench components and sequences were architected for strict reusability across hierarchy levels, enabling efficient regression testing within the constraints of wafer-scale complexity.

This paper details the verification methodology and architecture developed for the MOSAIX ASIC. Our primary contributions include:
\begin{itemize}
\item Hierarchical UVM Architecture: a scalable testbench framework capable of validating both isolated sensor tiles and the full wafer-scale integration without code duplication.
\item Test reusability: tests developed to be reused at different levels of abstraction: RTL, triplicated RTL and Gate-Level Netlist (GLN).
\item Retrospective Error Analysis: analyzing the GitLab database to track error trends over time, categorize issues by hierarchical block, and pinpoint the most frequent types of issues.
\end{itemize}

\section{MOSAIX ASIC}\label{sec:mosaix}

The MOSAIX ASIC was designed to address the stringent requirements of the ALICE ITS3 detector upgrade~\cite{tdr:alice:its3}.

\begin{figure}[h]
\caption{The MOSAIX ASIC block diagram: Left EndCap (LEC), Right EndCap (REC) and 12 Repeated Sensor Units (RSUs), each containing 12 tiles.}
\centering
\includegraphics[width=1\textwidth]{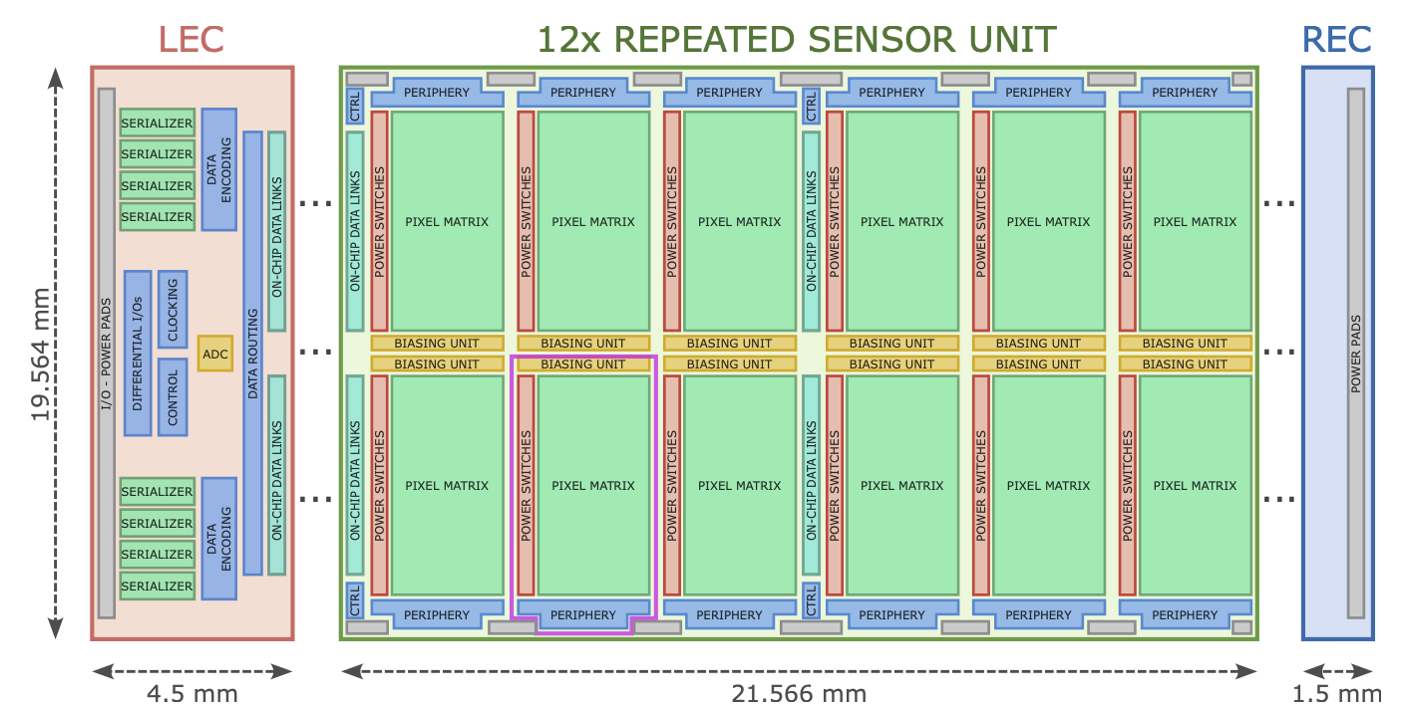}\label{fig:mosaix}
\end{figure}

To ensure operational yield, despite potential manufacturing defects, the MOSAIX ASIC, represented in Figure~\ref{fig:mosaix}, employs a highly modular architecture consisting of 12 Repeated Sensor Units (RSUs), aggregating a total of 144 independent pixel matrices tiles, where each tile integrates $444 \times 156$ pixels with a pitch of $22.8 \times 20.8\,\mu\text{m}^2$ and features an isolated power supply and data domain. The tiles can be selectively deactivated via on-chip switches in the event of manufacturing defects without disrupting the remainder of the wafer. 

Due to tight mechanical and material budget constraints, external service connections are restricted exclusively to the short peripheral edges of the chip via the endcaps. While global power is distributed symmetrically from both the Left EndCap (LEC) and the Right EndCap (REC) across the $26.66\,\text{cm}$ length of the wafer, the LEC serves as the exclusive hub for control input/output and data aggregation. The LEC gathers the parallel data links coming from all 144 internal tiles and streams them off-detector through high-speed transmitters continuously transmitting over $30\,\text{Gb/s}$ of particle collision data via integrated high-speed transmitters.

As the ITS3 layers operate in proximity to the beam pipe, relying solely on light air cooling, thermal management is critical. Therefore, the MOSAIX architecture is engineered to operate under a strict power density limit of less than $40\,\text{mW/cm}^2$ during data taking.

The ASIC implements selective Triple-Modular Redundancy (TMR) to increase the SEE tolerance: the control logic is completely triplicated (combinatorial and sequential logic, resets, clocks, voters), whereas the data path is not triplicated.

\section{Verification Strategy}\label{sec:vrf}
The verification strategy implements a coverage-driven constrained-random verification based on the Universal Verification Methodology (UVM)~\cite{ieee:uvm}.
Multiple verification components, developed for other ASICs in the High-Energy Physics community, were reused~\cite {chips:vip}.
The DUT was verified with the same verification framework at different levels of abstraction: RTL, triplicated RTL, and Gate-Level Netlist (GLN), with annotated delays.

The pixel matrix tile environment has been used to perform tile readout performance tests at different output frequencies, operating modes and occupancies. At MOSAIX scale, only an end-to-end data integrity test has been performed. 

Moreover, some tests could be performed only at the full wafer-scale level, such as Pseudo-Random Bit Sequence (PRBS) tests with PRBS generators inside each sensor tile and PRBS receivers in LEC.  

All hierarchies were also used to run extensive regressions with SEU injection at RTL level and GLN level.

\begin{figure}[h]
\caption{Example of verification plan with annotated coverage for features verified at sensor tile level. Results are from RTL regressions.}
\centering
\includegraphics[width=0.70\textwidth]{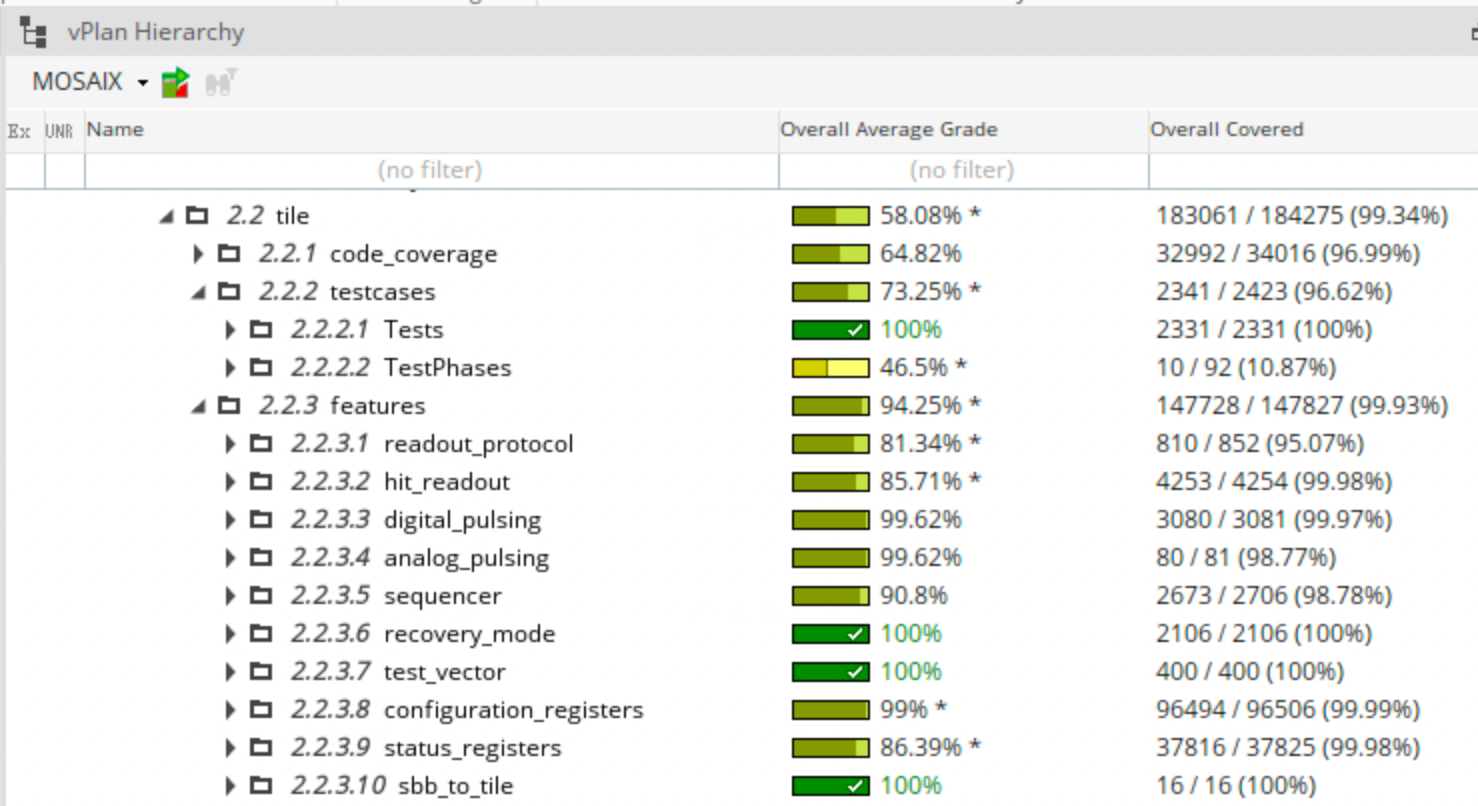}\label{fig:vplan}
\end{figure}

The verification strategy followed a clear verification plan devised before developing the test sequences. An example of the verification plan, with annotated coverage, is depicted in Figure~\ref{fig:vplan}.

Different kinds of coverage can be mapped into the verification plan. In our case, we used functional and formal coverage, as well as code coverage and designers' assertions coverage.

The environment and test sequence development tried to achieve $100\%$ coverage for all the features in the verification plan.

The final signoff could be achieved in around 200 days, normalized to 64 single-core simulations.

\section{Retrospective analysis of found errors}\label{sec:tdc}
\begin{figure}[h]
\caption{Graph showing the number of errors over time found during RTL (blue) and in GLN (red) during project development.}
\centering
\includegraphics[width=0.90\textwidth]{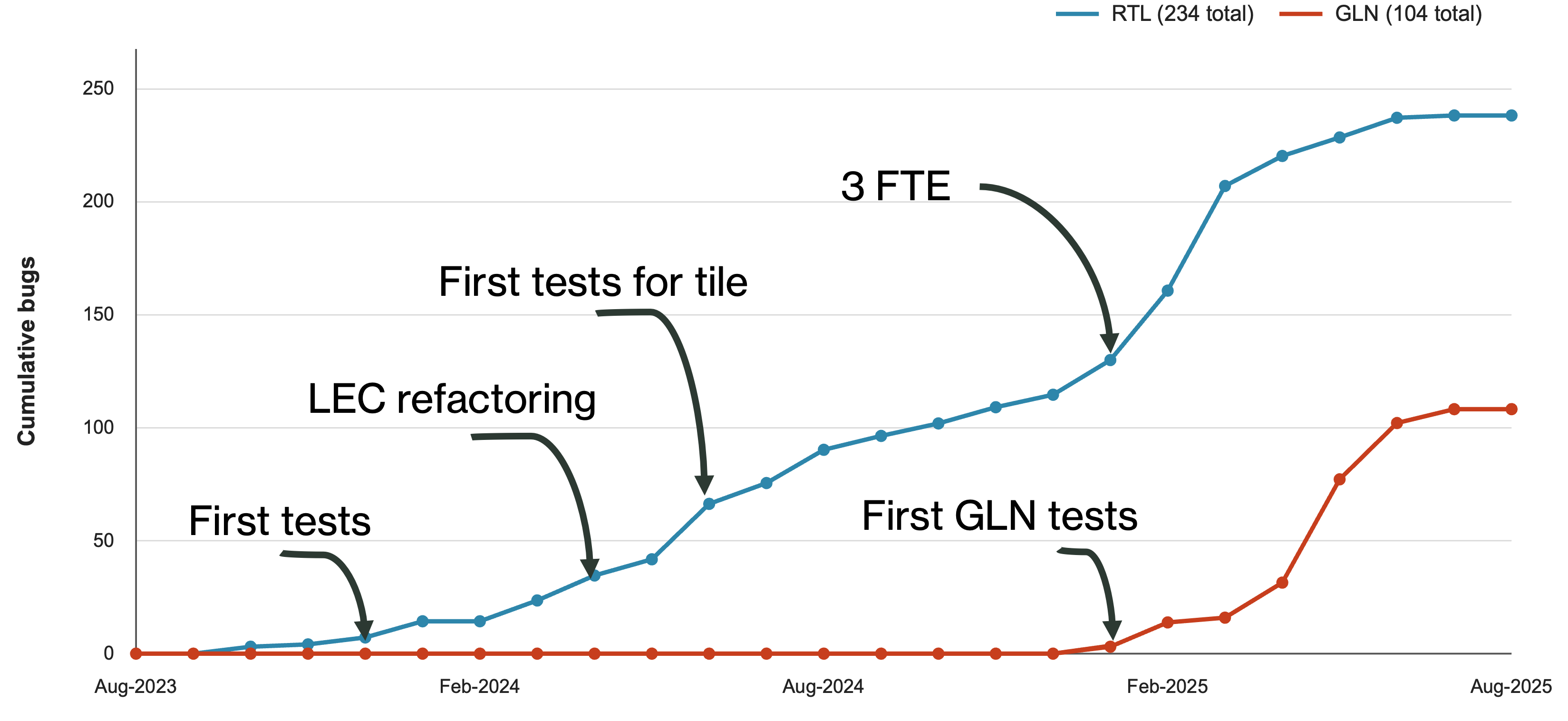}\label{fig:errors_history}
\end{figure}

By analyzing the GitLab database after the submission, the number of errors found during RTL and GLN simulations was extracted. Figure~\ref{fig:errors_history} reports the total number of errors found during RTL and GLN simulations. The graph only shows data from August 2023, which is when the verification team was brought on board with an average of 1.5 FTE.

Following a few months of initial development, the first errors emerged at the RTL level. Subsequent modifications to the Left End Cap block altered feature behavior and required validation. As the tile testbench matured, it uncovered a wave of errors linked to undefined or poorly documented specifications, driving a steady increase in error counts from August 2024 to February 2025. By February 2025, the team went from 1.5 FTE to 3 FTE. At this point, with other kinds of errors, Single Event Upset (SEU) simulations also started and uncovered a large fraction of issues that required fixing. At the same time, the first gate-level netlist (GLN) was generated. After baseline checks at the typical corner to prove the environment was ready to run GLN, minimum and maximum corners were exercised and exposed timing violations. It took a total of seven months from the arrival of the first GLN to achieve ASIC submission.

\begin{figure}[h]
\caption{Graph showing the number of errors found in different design blocks at RTL (blue) and GLN (red) level.}
\centering
\includegraphics[width=0.90\textwidth]{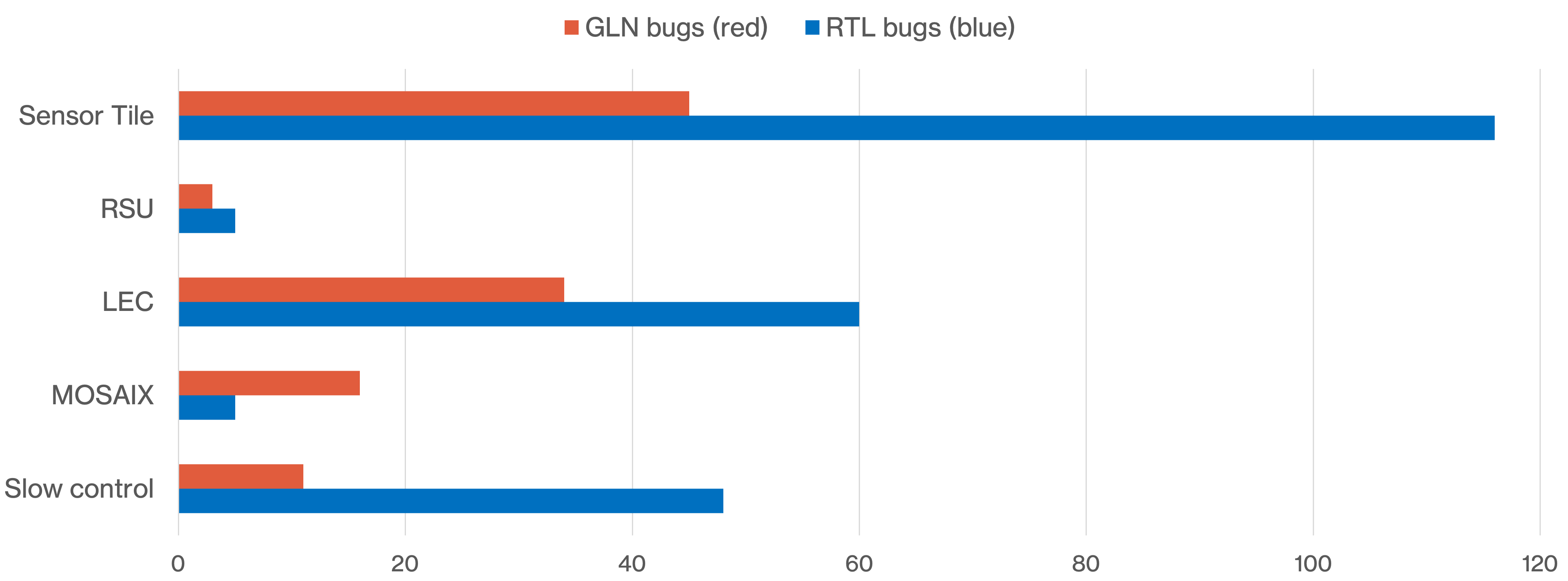}\label{fig:errors_by_block}
\end{figure}

Figure~\ref{fig:errors_by_block} illustrates the distribution of errors across the four verification hierarchies, in addition to a shared "slow control" category common to all hierarchies.
As illustrated, the sensor tile exhibited the highest error count in both the RTL and GLN levels of abstraction. This concentration of issues stems from the block's intricate, interacting FSMs and complex features, which significantly complicated forecasting the verification closure.

\begin{figure}[h]
\caption{Pie chart reporting the errors found by level of abstraction (RTL, triplicated RTL and GLN) plus a fraction due to verification.}
\centering
\includegraphics[width=0.90\textwidth]{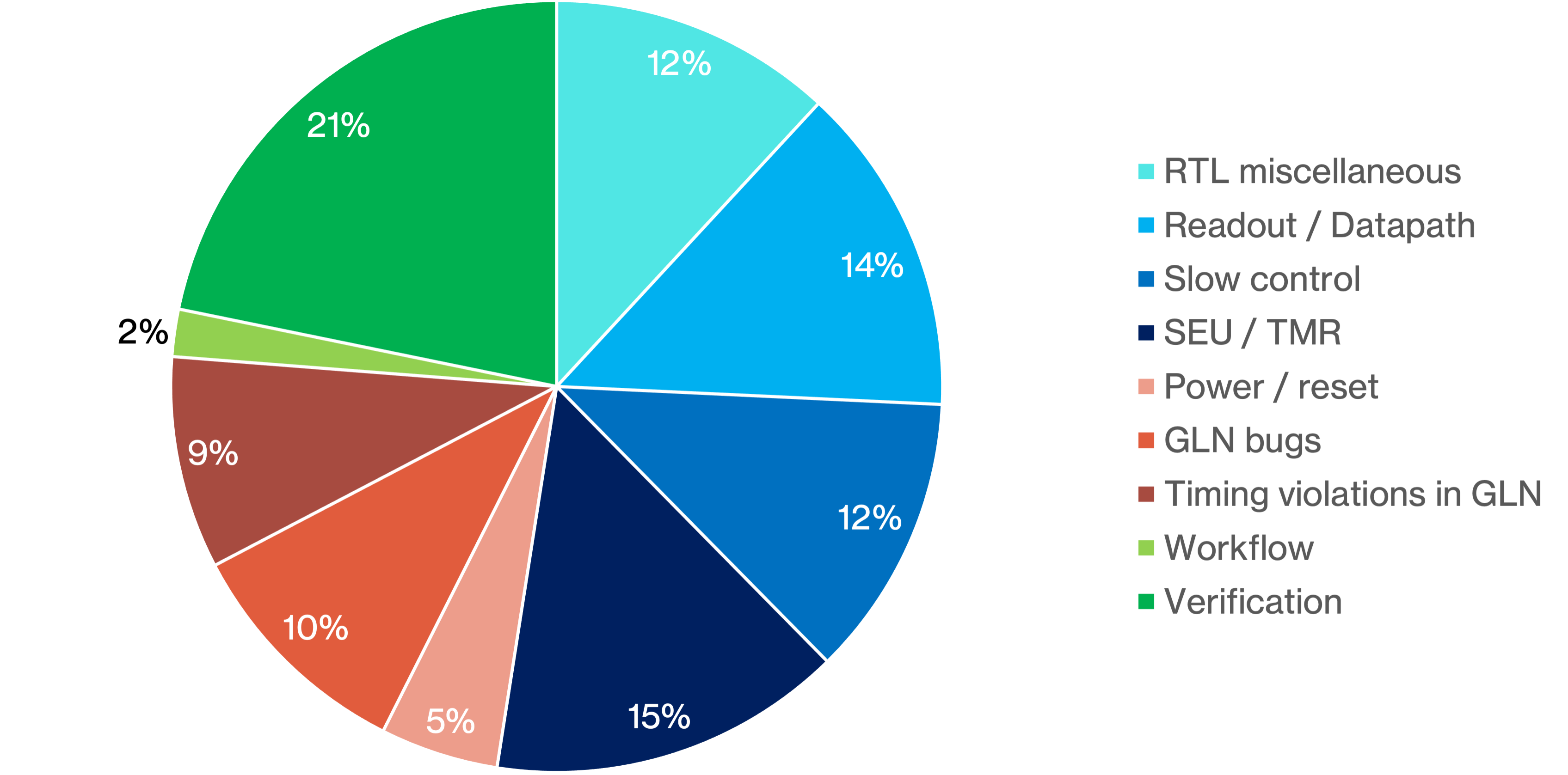}\label{fig:errors_kinds}
\end{figure}

Another valuable perspective categorizes the identified errors by abstraction level: RTL, triplicated RTL, and GLN. The distribution illustrated in Figure~\ref{fig:errors_kinds} highlights the necessity of a comprehensive verification flow in HEP projects, spanning from initial RTL, through triplicated RTL, to final GLN. Evaluated clockwise, the first three light blue sectors represent RTL-level errors (38\%). The RTL miscellaneous category predominantly covers semantic errors detected through formal verification. The readout category encompasses control logic issues—such as ghost packets, lost packets, and repeated pixels driven by readout FSMs, as well as FIFO boundary corner cases that caused unexpected words and sporadic lost hits.
SEU evaluation yielded (15\%) of the total error count. Given their critical impact on the target application and the subsequent requirement for RTL remediation, these findings underscore the importance of initiating SEU campaigns early in the design cycle.
The orange and brown sectors indicate that (24\%) of errors were isolated at the GLN stage. While a small subset (5\%) emerged during power regressions, the majority originated from timing constraints, including incorrectly implemented CDCs, false paths, and multi-cycle paths. Finally, the green sector (23\%) denotes verification overhead induced by documentation ambiguity or omitted specifications, which ultimately forced the redesign of scoreboards and monitors to properly capture complex reset behavior during active data acquisition.

\section{Conclusions}\label{sec:end}
This work presents the verification strategy of the MOSAIX ASIC, emphasizing the critical importance of a solid verification plan prior to environment development. Furthermore, hierarchical environment and sequence reuse, and test reuse are demonstrated to be fundamental to achieving verification closure within a reasonable timeframe (approximately 200 days normalized to 64 single-core simulations).

Additionally, a retrospective look at the detected errors highlights that short-cutting the verification process to save time risks compromising design quality, leading to escaped errors that could result in costly late redesign.

The success of this verification framework is demonstrated by the real-world performance of the MOSAIX prototype, which was taped out in August 2025 and received in February 2026. In-depth hardware evaluation proved that the ASIC functions exactly as specified, directly validating the pre-silicon testbench architecture. This robust foundation has accelerated the design cycle for the final MOSAIX, which is currently being submitted for fabrication.



\newpage
\begin{thebibliography}{99}

\bibitem{tdr:alice:its3}
The ALICE collaboration,
\emph{{Technical Design report for the ALICE Inner Tracking System 3 - ITS3; A bent wafer-scale monolithic pixel detector}},
\emph{ALICE-TDR-021} \emph{CERN-LHCC-2024-003} (2024),

\bibitem{mosaix:design}
Leitao, P.
\emph{{Development of the MOSAIX chip for the ALICE ITS3 upgrade}},
\emph{{2025 JINST 20 C06001}},
DOI: {10.1088/1748-0221/20/06/C06001}

\bibitem{ieee:uvm}
\emph{{IEEE Standard for Universal Verification Methodology Language Reference Manual}},
\emph{{IEEE Std 1800.2-2020}} (2020),
DOI: {10.1109/IEEESTD.2020.9195920}

\bibitem{chips:vip}
M. Lupi et al.,
\emph{Reusable verification components for High-Energy Physics readout ASICs},
\emph{{2025 JINST 20 C02002}}
DOI: {10.1088/1748-0221/20/02/C02002}

\end{thebibliography}
\end{document}